# High-angular resolution electron backscatter diffraction as a new tool for mapping lattice distortion in geological minerals

D. Wallis[1], L. N. Hansen[2], T. B. Britton[3], and A. J. Wilkinson[4]

[1]Department of Earth Sciences, Utrecht University, Utrecht, The Netherlands.

[2]Department of Earth Sciences, University of Oxford, Oxford, U.K.

[3]Department of Materials, Imperial College London, London, U.K.

[4]Department of Materials, University of Oxford, Oxford, U.K.

Corresponding author: David Wallis (d.wallis@uu.nl)

**Key Points:**

- HR-EBSD uses cross correlation of diffraction patterns to map lattice distortion.

- Rotations and strains can be used to calculate GND densities and residual stresses, respectively.

- Recent developments in data analysis make HR-EBSD suitable for a wide range of rocks.

**Abstract**

Analysis of distortions of the crystal lattice within individual mineral grains is central to the investigation of microscale processes that control and record tectonic events. These distortions are generally combinations of lattice rotations and elastic strains, but a lack of suitable observational techniques has prevented these components being mapped simultaneously and routinely in earth science laboratories. However, the technique of high-angular resolution electron backscatter diffraction (HR-EBSD) provides the opportunity to simultaneously map lattice rotations and elastic strains with exceptional precision, on the order of 0.01° for rotations and $10^{-4}$ in strain, using a scanning electron microscope. Importantly, these rotations and lattice strains relate to densities of geometrically necessary dislocations and residual stresses. Recent works have begun to apply and adapt HR-EBSD to geological minerals, highlighting the potential of the technique to provide new insights into the microphysics of rock deformation. Therefore, the purpose of this review is to provide a summary of the technique, to identify caveats and targets for further development, and to suggest areas where it offers potential for major advances. In particular, HR-EBSD is well suited to characterising the roles of different dislocation types during crystal plastic deformation and to mapping heterogeneous internal stress fields associated with specific deformation mechanisms/microstructures or changes in temperature, confining pressure, or macroscopic deviatoric stress. These capabilities make HR-EBSD a particularly powerful new technique for analysing the microstructures of deformed geological materials.

**1. Introduction**

The rates and styles of geodynamic processes on rocky planets emerge from a complex ensemble of underlying processes operating at scales down to the crystal lattices of their constituent minerals.

Deciphering this emergent behaviour is one of the principal aims of the solid-earth geosciences. Analysis of rock microstructures is central to this effort in two respects. First, the microstructures of crystalline materials exert some of the key controls on their mechanical properties. Second, microstructures in deformed materials provide an invaluable record of the processes that operated during, and after, deformation. As such, microstructural data provide a fundamental basis for developing models of deformation behaviour and testing their applicability to both experimental and natural systems. Therefore, the development and refinement of techniques that provide microstructural data has been one of the main drivers of recent advances in the fields of experimental rock deformation, structural geology, and tectonics.

A key technique for microstructural analysis of geological materials is electron backscatter diffraction (EBSD) (Dingley, 1984; Wilkinson and Hirsch, 1997; Prior et al., 1999, 2009). EBSD is based on analysis of diffraction patterns acquired in a scanning electron microscope (SEM) and provides a diverse range of microstructural data relevant to geological questions. Some EBSD datasets characterise the microstructures of aggregates of grains. Examples include the distributions of phases, lattice orientations, intergranular misorientations, and grain sizes and shapes (Prior et al., 1999, 2009). Other datasets characterise intragranular microstructures, particularly intragranular lattice misorientations, which are used to constrain the types of dislocations present (Trimby et al., 1998; Bestmann and Prior, 2003; Lloyd, 2004). Due to the relative ease with which these rich microstructural data can be acquired, EBSD analysis has become routine in many earth-science laboratories and, whilst these data are employed in a wide range of geoscience subdisciplines, they have become particularly central to the study of deformed rocks (Prior et al., 2009; Parsons et al., 2016; Cross et al., 2017; Tasaka et al., 2017; Tommasi et al., 2017; Weikusat et al., 2017; Ceccato et al., 2018; Wallis et al., 2018).

Despite its diverse capabilities, conventional EBSD has some key limitations in its ability to characterise subtle intragranular lattice distortions that provide important records of deformation processes. Crystal orientations are most commonly determined by indexing the Hough transforms of diffraction patterns to a database of crystal structures (Wright and Adams, 1992; Adams et al., 1993). However, locating the peaks in Hough space limits the precision in measurements of (mis)orientation to the order of ~0.1° (Humphreys et al., 1999). Whilst this angular resolution is sufficient for many purposes, subtle but potentially valuable details of the substructure can be obscured. Furthermore, the precision in misorientation *axes* decreases with decreasing misorientation *angle* to the extent that measured misorientation axes can deviate from their true values by tens of degrees for misorientation angles on the order of 1° (Prior, 1999; Wilkinson, 2001). The presence of geometrically necessary dislocations (GNDs) can result in lattice curvature over a specified length-scale, as these dislocations have a net Burgers vector that does not cancel out. Therefore, the precision in measured misorientation angles and axes limits respectively the densities and types of GNDs that can be resolved (Wallis et al., 2016). Moreover, the Hough transform-based indexing approach does not allow recovery of information on the variations in elastic strain state of the crystal. However, elastic strains and their associated stresses can exert important controls on deformation processes and other microstructural changes.

Developments in the materials sciences over the last decade or so have largely overcome these limitations by developing an alternative data processing approach, termed high-angular resolution electron backscatter diffraction (HR-EBSD). HR-EBSD is based on cross correlation of multiple regions of interest between diffraction patterns to measure the deformation gradient tensor (Wilkinson, 1996; Wilkinson et al., 2006a; Britton and Wilkinson, 2011). This tensor can be decomposed into lattice rotations and elastic strains, both of which can be measured to a precision of $< 10^{-4}$. This precision corresponds to the order of 0.01° for lattice rotations (Wilkinson, 1996; Wilkinson et al., 2006a). Using this approach, precision in the axes of small misorientations is also dramatically improved over traditional indexing of EBSD patterns (Wilkinson, 2001). The improved precision in misorientation angles and axes translates into improved precision in estimates of the corresponding densities and types of GNDs, respectively (Jiang et al., 2013a; Ruggles et al., 2016a; Wallis et al., 2016). Moreover, the elastic strain data can be converted into maps of residual stresses retained within the microstructure (Karamched and Wilkinson, 2011; Britton and Wilkinson, 2012a; Jiang et al., 2013b).

The ability to precisely map intragranular lattice rotations, GND densities, elastic strains, and residual stresses using EBSD data collected in a standard SEM has led to a wealth of developments in the materials sciences over the past decade. Examples include analyses of distributions of GNDs and elastic strain and residual stress heterogeneity in deformed metals (Wilkinson and Randman, 2010; Jiang et al., 2015a), alloys (Britton et al., 2010; Littlewood et al., 2011; Jiang et al., 2016), semiconductors (Vilalta-Clemente et al., 2017) and ceramics (Villanova et al., 2012), along with characterisation of specific processes, such as interactions between dislocations and grain boundaries (Britton and Wilkinson, 2012a), amongst many others.

The success of HR-EBSD in the materials sciences alludes to the potential of the technique to offer new insights into the microstructures, deformation processes, and mechanical properties of analogous geological materials. Schäffer et al. (2014) provided an early application of HR-EBSD to geological materials by mapping apparent strains resulting from changes in lattice parameters caused by solid solution in alkali feldspars. Over the past few years, we have undertaken initial HR-EBSD analyses of deformed geological minerals. Examples include mapping GNDs and residual stress heterogeneity in single crystals of olivine (Wallis et al., 2016, 2017a; Kumamoto et al., 2017) and mapping GNDs in aggregates of olivine (Boneh et al., 2017; Kumamoto et al., 2017; Qi et al., 2018) and quartz (Wallis et al., 2017b). These examples demonstrate the great potential of the technique and highlight some subtle but important considerations for analysis of geological materials in particular. Therefore, it is timely to provide a review of the application of HR-EBSD in the earth sciences. We begin by providing summaries of the technique and practical aspects of its application, then highlight key points by providing illustrative examples. We finish by discussing the strengths and limitations of the technique and summarising potential research directions.

## 2. Technique development

HR-EBSD has developed gradually over the past 25 years, but only recently have developments made it suitable for wide-ranging application to the variety of typical rock microstructures. Early works recognised the potential of measuring small shifts of features within EBSD patterns to reveal small lattice rotations and elastic strains (Troost et al., 1993; Wilkinson, 1996, 2000, 2001). A major development was made by Wilkinson *et al*. (2006a), who presented a practical and mathematical framework for estimating eight degrees of freedom in the displacement gradient tensor, describing rotations and strains, from diffraction patterns obtained on megapixel charge-coupled device detectors. The final degree of freedom relates to the hydrostatic strain, which cannot be measured directly but can be determined by constraining the surface normal stress to zero (Wilkinson et al., 2006b). Subsequent methodological refinements have focussed on assessing and improving the accuracy and precision of strain measurement and extending the potential applications to more challenging microstructures (Maurice and Fortunier, 2008; Villert et al., 2009; Britton et al., 2010; Britton and Wilkinson, 2011, 2012b; Maurice et al., 2012; Britton et al., 2013a, 2013b; Wilkinson et al., 2014; Plancher et al., 2015; Tong et al., 2015). A recent advance that is particularly important for analysis of geological materials, which are commonly deformed to large plastic strains, has been the development of routines for mapping elastic strains in the presence of lattice rotations of several degrees, such as subgrain boundaries (Britton and Wilkinson, 2011, 2012b; Maurice et al., 2012). Below, we summarise the key elements of the technique that are necessary to appreciate its application to geological materials.

## 3. Principles of HR-EBSD

### *3.1. Measuring rotations, elastic strains, and residual stresses*

HR-EBSD data are derived from mapping small distortions within stored images of diffraction patterns. In this section, we summarise the method of mapping these distortions based on the work of Wilkinson et al. (2006a) and improved by Britton and Wilkinson (2011, 2012b). For HR-EBSD, the Hough transform is used only to determine the orientation of a single reference point in each grain. Consequently, the accuracy of *absolute* crystal orientations in HR-EBSD datasets is the same as the original EBSD data, on the order of a few degrees and limited by specimen alignment and geometric distortions (Nolze, 2007). However, HR-EBSD processing improves the precision in relative orientations, i.e., misorientations, between a reference point in a grain of interest and all other points in that grain. An array of small, typically 256 x 256 pixel, regions of interest (ROI) are extracted from the same positions within each ~1000 x 1000 pixel diffraction pattern. A minimum of 4 dispersed ROI are required from each diffraction pattern, but typically 20 or more are employed to over-determine the deformation gradient tensor. Misorientations between the reference point and all other points in the grain are determined by cross correlating each ROI from each diffraction pattern with the corresponding ROI from the reference pattern. For computational speed, the fast Fourier transforms of the ROI are computed, and the cross correlation is performed in Fourier space. This procedure has the added benefit that bandpass filters can be easily applied to reduce high-frequency noise and long-wavelength intensity gradients within the ROIs. The position of the peak in the cross-correlation function gives the translation that best aligns the ROIs from the reference and test patterns, that is, it determines how the position of the ROI is shifted between one pattern and the other. To improve the precision in the shift measurement, the peak in the cross-correlation function is interpolated to estimate its position to ± 0.05 pixels, which provides a precision in strain of $10^{-4}$ when working with megapixel diffraction patterns (Villert et al., 2009).

The geometry of a displaced ROI in a diffraction pattern is illustrated schematically in Figure 1. The position of the centre of a ROI in the reference pattern is given by vector **r**. In a test pattern, the feature found in the reference pattern at **r** would be instead projected to position **r'**. Only the component of this displacement that lies within the plane of the phosphor screen is detected as a shift, **Q**, in the position of the ROI. The component of **r'** extending out of the plane of the phosphor screen, Δ**r'**, is not detected.

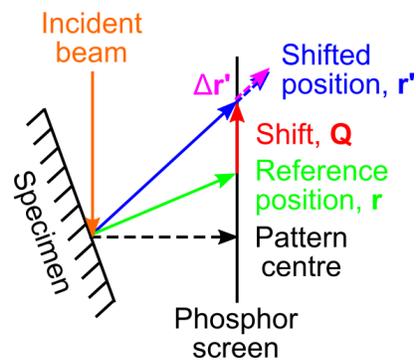

**Figure 1**. Schematic of the geometry involved in describing a shift, **Q**, of a point within a diffraction pattern from a reference position, **r**, to a shifted position, **r'**. The component of **r'**

extending out of the plane of the phosphor screen, Δr', cannot be directly detected from a shift within the plane of the phosphor screen. After Britton and Wilkinson (2012b).

The mapping of **r** to **r'** is described by

$$\mathbf{r'} = \boldsymbol{\beta}\mathbf{r}, \tag{1}$$

where $\boldsymbol{\beta}$ is the deformation gradient tensor. An example of this mapping is presented in Figure 2. $\boldsymbol{\beta}$ is determined by fitting the shifts in the ROIs. When the elastic strains and lattice rotations are small, $\boldsymbol{\beta}$ can be additively decomposed as

$$\boldsymbol{\beta} = \boldsymbol{\varepsilon} + \boldsymbol{\omega} + \mathbf{I}, \tag{2}$$

where **I** is the identity matrix and $\boldsymbol{\varepsilon}$ and $\boldsymbol{\omega}$ are the infinitesimal strains and rotations, respectively, given by

$$\boldsymbol{\varepsilon} = \frac{1}{2}(\boldsymbol{\beta} + \boldsymbol{\beta}^T) - \mathbf{I}, \tag{3}$$

where $^T$ indicates the transpose, and

$$\boldsymbol{\omega} = \frac{1}{2}(\boldsymbol{\beta} - \boldsymbol{\beta}^T) - \mathbf{I}. \tag{4}$$

The stress normal to the section is assumed to be zero, and this provides constraint to solve for the hydrostatic strain.

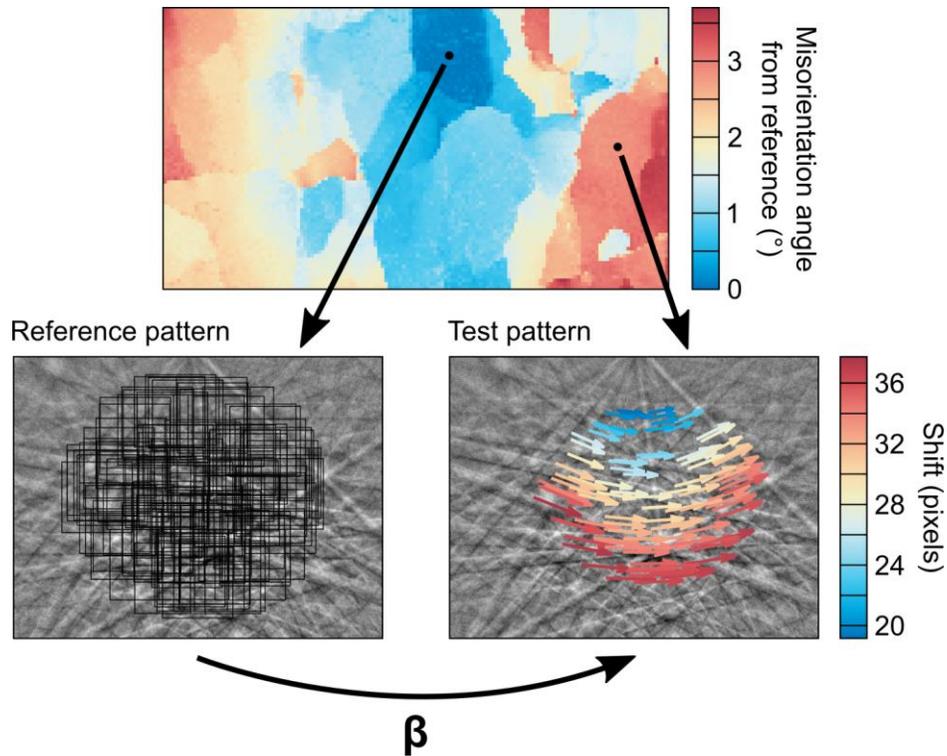

**Figure 2**: Example of shifts in regions of interest between the reference pattern and a test pattern in a different subgrain in quartz (sample P12/058 of Parsons et al. (2016) and Wallis et al. (2017b)). 100 regions of interest, each 256 x 256 pixels, are marked by black boxes on the reference pattern. Exaggerated shifts in these regions of interest are indicated by arrows on the test pattern. The true magnitude of the shifts is indicated by the colour scale. The shifts in the regions of interest are described by the deformation gradient tensor, **β**.

The cross-correlation basis of HR-EBSD places limits on the maximum misorientation angle between pairs of patterns that can be analysed. The method originally presented by Wilkinson et al. (2006a) can measure lattice rotations up to misorientation angles of approximately 8°, beyond which the patterns become too dissimilar and distorted for accurate measurements (Britton and Wilkinson, 2011). Elastic strain measurements are limited to an even more restricted range of misorientation angles of approximately 1°, beyond which the large shifts due to the rotations swamp the much smaller signal from elastic strains (Britton and Wilkinson, 2012b). Beyond this angular range, errors introduced to the strain estimate by the large rotations commonly result in spuriously large 'phantom strains' (Britton and Wilkinson, 2012b).

To extend the range of misorientation angles over which rotations and strains can be measured, Britton and Wilkinson (2012b) proposed a more advanced three-step approach. First, an initial pass of cross correlation is employed to obtain an infinitesimal rotation matrix based on the procedure outlined above, whilst elastic strains are initially ignored. Second, the infinitesimal rotations are used to estimate a finite rotation matrix, $\mathbf{\Omega}$, by

$$\mathbf{\Omega} = \begin{pmatrix} \cos \omega_{12} & \sin \omega_{12} & 0 \\ -\sin \omega_{12} & \cos \omega_{12} & 0 \\ 0 & 0 & 1 \end{pmatrix} \times \begin{pmatrix} 1 & 0 & 0 \\ 0 & \cos \omega_{23} & \sin \omega_{23} \\ 0 & -\sin \omega_{23} & \cos \omega_{23} \end{pmatrix} \times \begin{pmatrix} \cos \omega_{31} & 0 & -\sin \omega_{31} \\ 0 & 1 & 0 \\ \sin \omega_{31} & 0 & \cos \omega_{31} \end{pmatrix}. \quad (7)$$

This finite rotation matrix is used to make a virtual rotation of the test pattern back into an orientation similar to that of the reference pattern. In this step, the intensities in the measured test pattern are interpolated and remapped to generate a rotated test pattern. Third, a second pass of cross correlation is employed to measure elastic strains from the rotated test pattern (which is now much more similar to the reference pattern), along with any small corrections to the rotations. A similar methodology has also been presented by Maurice *et al*. (2012). This remapping approach extends the angular range over which both rotations and strains can be measured to ~11°. This procedure dramatically extends the range of materials in which elastic strains can be measured and, importantly, allows elastic strain measurements in materials containing large plastic strain gradients manifesting as intragranular misorientations of several degrees. Based on the elastic strain measurements, it is straightforward to calculate the corresponding residual stresses based on

Hooke's law and the elastic stiffness tensor of the material (Karamched and Wilkinson, 2011; Britton and Wilkinson, 2012a).

As HR-EBSD datasets are typically maps containing large numbers of individual measurements it has been important to establish some data quality metrics to assure that points with poorer quality data (perhaps associated with poor condition of the specimen surface) can be readily identified and used to exclude such points from further analysis. Two main measures are used (Britton and Wilkinson, 2011). The first metric is the geometric mean of the cross-correlation peak heights determined for each region of interest. If the pattern matching and registration has worked well, values close to unity are obtained, but the value falls if, for example, the pattern is from a different grain or if dust or a surface pore lead to partial shadowing of the pattern. The second metric is the (weighted) mean angular error, which assesses the fit of the solution for the deformation gradient tensor, $\boldsymbol{\beta}$, to the measured shifts in the ROIs. The mean angular error is the arithmetic mean of the difference in angular shift predicted at the centre of each ROI and that actually measured. Therefore, strain and rotation measurements that are smaller than the mean angular error should be treated with caution.

### *3.2. Estimating densities and types of geometrically necessary dislocations*

Curvature of the crystal lattice results from the presence of geometrically necessary dislocations. Analysis of GNDs can be conducted through the 'dislocation tensor', $\boldsymbol{\alpha}$, using Nye-Kröner analysis (Nye, 1953; Kröner, 1958). The presence of dislocations introduces spatial gradients, in the directions $x_i$, of lattice orientation (measured as rotations) and elastic strain, which contribute to the components $\alpha_{ij}$ of $\boldsymbol{\alpha}$ by

$$\alpha_{ij} = \begin{bmatrix} \dfrac{\partial \omega_{12}}{\partial x_3} - \dfrac{\partial \omega_{31}}{\partial x_2} & \dfrac{\partial \omega_{13}}{\partial x_1} & \dfrac{\partial \omega_{21}}{\partial x_1} \\ \dfrac{\partial \omega_{32}}{\partial x_2} & \dfrac{\partial \omega_{23}}{\partial x_1} - \dfrac{\partial \omega_{21}}{\partial x_3} & \dfrac{\partial \omega_{21}}{\partial x_2} \\ \dfrac{\partial \omega_{32}}{\partial x_3} & \dfrac{\partial \omega_{13}}{\partial x_3} & \dfrac{\partial \omega_{31}}{\partial x_2} - \dfrac{\partial \omega_{32}}{\partial x_1} \end{bmatrix} \qquad (8)$$

$$+ \begin{bmatrix} \dfrac{\partial \varepsilon_{12}}{\partial x_3} - \dfrac{\partial \varepsilon_{13}}{\partial x_2} & \dfrac{\partial \varepsilon_{13}}{\partial x_1} - \dfrac{\partial \varepsilon_{11}}{\partial x_3} & \dfrac{\partial \varepsilon_{11}}{\partial x_2} - \dfrac{\partial \varepsilon_{12}}{\partial x_1} \\ \dfrac{\partial \varepsilon_{22}}{\partial x_3} - \dfrac{\partial \varepsilon_{23}}{\partial x_2} & \dfrac{\partial \varepsilon_{23}}{\partial x_1} - \dfrac{\partial \varepsilon_{21}}{\partial x_1} & \dfrac{\partial \varepsilon_{21}}{\partial x_2} - \dfrac{\partial \varepsilon_{22}}{\partial x_1} \\ \dfrac{\partial \varepsilon_{32}}{\partial x_3} - \dfrac{\partial \varepsilon_{33}}{\partial x_2} & \dfrac{\partial \varepsilon_{33}}{\partial x_1} - \dfrac{\partial \varepsilon_{31}}{\partial x_3} & \dfrac{\partial \varepsilon_{31}}{\partial x_2} - \dfrac{\partial \varepsilon_{32}}{\partial x_1} \end{bmatrix}.$$

The components of $\alpha_{ij}$ relate to the densities, $\rho^s$, of $s_{max}$ different types of dislocation, with Burgers vectors $\mathbf{b}^s$ and line directions $\mathbf{l}^s$, through

$$\alpha_{ij} = \sum_{s=1}^{s_{max}} \rho^s b_i^s l_j^s. \qquad (9)$$

It is clear from Equations 8 and 9 that different types of dislocation generate different rotation and strain gradients and hence contribute differently to the components of $\alpha_{ij}$. For instance, screw dislocations (with Burgers vectors parallel to their line directions) contribute to the diagonal components, whereas edge dislocations (with Burgers vectors perpendicular to their line directions) contribute to the off-diagonal components. Dislocations of mixed character are represented by densities of their endmember edge and screw components. Rotation and strain gradients in the direction normal to the specimen surface (i.e., in the $x_3$ direction) cannot be measured from a two-dimensional EBSD map. The absence of this information leaves only the $\alpha_{i3}$ terms fully determined. However, often the rotation gradients are larger than the elastic strain gradients (an assessment that can be made from HR-EBSD data but not from conventional EBSD data), in which case, the elastic strain gradients can be neglected entirely or only the measurable terms included in the analysis (Wilkinson and Randman, 2010). In this case, five components of

$\alpha_{ij}$ ($\alpha_{12}$, $\alpha_{13}$, $\alpha_{21}$, $\alpha_{23}$, and $\alpha_{33}$) can be determined directly, along with the difference between two of the remaining components, i.e., $\alpha_{11}-\alpha_{22}$ (Pantleon, 2008).

For simple cubic crystals with **b**$^s$ and **l**$^s$ of nine (or fewer) dislocation types constrained to lie along the cube axes, as originally considered by Nye (1953), Equation 9 provides an intuitive and unambiguous relationship between the lattice curvature and dislocation content (Arsenlis and Parks, 1999; Sun et al., 2000; Wilkinson and Randman, 2010). However, to analyse more complex crystal structures with more numerous possible dislocation types, a more general approach is required. The problem of estimating the densities of each type of GND from the available components of $\alpha_{ij}$ can be set out as

$$\mathbf{A}\boldsymbol{\rho} = \boldsymbol{\lambda}, \tag{10}$$

where $\boldsymbol{\rho}$ is a vector of the densities of the $s_{max}$ dislocation types and $\boldsymbol{\lambda}$ is a vector containing the measurable components of lattice curvature, of which there are six, corresponding to components of $\alpha_{ij}$ (Pantleon, 2008). **A** is a $6 \times s_{max}$ matrix in which each column contains the dyadic of the Burgers vector and unit line direction of the $s$th dislocation type (Arsenlis and Parks, 1999; Britton and Wilkinson, 2012a). Equation 10 can be solved using the right Moore-Penrose inverse,

$$\boldsymbol{\rho} = \mathbf{A}^{\mathrm{T}}(\mathbf{A}\mathbf{A}^{\mathrm{T}})^{-1}\boldsymbol{\lambda} \tag{11}$$

(Arsenlis and Parks, 1999; Wilkinson and Randman, 2010). Equation 11 can be used to directly calculate the best-fit values of $\boldsymbol{\rho}$ since it inherently solves Equation 10 in a least-squares sense by minimising the $L_2$-norm of the dislocation densities,

$$L_2 = \left[\sum_{s=1}^{s_{max}} [\rho^s]^2\right]^{1/2} \tag{12}$$

(Dunne et al., 2012). This approach yields a unique solution for crystal structures in which the analysis can reasonably be limited to consideration of six dislocation types or fewer, and has been employed to estimate GND densities in olivine deformed at high temperature (Wallis et al., 2016, 2017a; Boneh et al., 2017; Kumamoto et al., 2017; Qi et al., 2018). However, for crystal structures with more than six dislocation types, there are typically many combinations of dislocation types and densities that are geometrically capable of generating measured lattice curvature, and the problem of solving Equation 10 is underconstrained (i.e., **A** has more columns than rows). In such cases, an additional constraint must be employed to select an optimal solution. One such approach that has been applied to cubic and hexagonal metals (Wilkinson and Randman, 2010; Britton and Wilkinson, 2012a; Jiang et al., 2013c) is to weight the dislocation densities by their line energy, $E^s$, in the minimisation of $L_1$ in

$$L_1 = \sum_{s=1}^{s_{max}} |\rho^s E^s|. \tag{13}$$

The energies of edge and screw dislocations, $E_{edge}$ and $E_{screw}$ respectively, used in the $L_1$ minimisation scheme are in the ratio

$$\frac{E_{edge}}{E_{screw}} = \frac{1}{1-\nu}, \tag{14}$$

where $\nu$ is the Poisson's ratio (Wilkinson and Randman, 2010). This approach has been applied to quartz, considering 19 dislocation types grouped into six families (Wallis et al., 2017b). As calculation of the energies of individual dislocations and the combined energies of groups of dislocations is complex (Popov and Kröner, 2001; Mesarovic et al., 2015), the above approach is a simplification to provide a tractable solution. One additional effect is the impact of elastic anisotropy on dislocation energy, which is on the order of 10% for common geological minerals (Heinisch et al., 1975) and therefore will often be of secondary importance in situations where the lattice curvature alone dictates orders of magnitude difference in the densities of different dislocation types. Another simplification is that the energies are based on those of single dislocations rather than considering the net effects of elastically interacting groups of dislocations. The energy of groups of dislocations is a difficult problem because it depends on the precise arrangement of the dislocations, but is the subject of ongoing work (Zheng et al., 2019).

The sensitivity of GND density estimates depends on three main factors, specifically, the precision ($\theta$) in lattice rotation measurements, the mapping step size ($d$), and the lattice orientation in the specimen reference frame. The first two factors control the precision of calculated orientation gradients and therefore the GND densities estimated from them. The precision of rotation measurements is controlled by a complex interplay of factors, including pattern size and quality along with data processing options, such as the size and number of ROIs (Wilkinson et al., 2006a; Britton and Wilkinson, 2011). As noted above, with optimal data acquisition and processing, precision on the order of $10^{-4}$ can be achieved in practice (Wilkinson et al., 2006a; Britton and Wilkinson, 2011). Subtle (monotonic) orientation gradients are easier to detect if the measurement points are further apart so that the orientation difference is greater. Therefore, large step sizes improve precision in measured orientation gradients, albeit at the expense of spatial resolution. A simple estimate of the minimum resolvable density ($\rho_{\min}$) of GNDs with a Burgers vector of magnitude $b$ can be made by

$$\rho_{\min} = \frac{\theta}{bd}. \tag{15}$$

The minimum resolvable GND density is referred to as the 'noise floor' because GND densities below this level are obscured by those calculated from noise in the rotation measurements (Jiang et al., 2013a; Wallis et al., 2016). The effect of lattice orientation on the noise floor in GND density estimates is more complex and has been addressed in detail by Wheeler et al. (2009) and Wallis et al. (2016). The key point is that, if the Burgers vector of an edge dislocation is orientated normal to the specimen surface, then that dislocation produces no orientation gradients detectable in the plane of observation. This effect has two main consequences. First, such dislocations cannot be detected by (mis)orientation data collected from that surface, providing one reason why GND density estimates provide a lower bound on the total dislocation density. Second, as these unfavourably oriented dislocations each produce little apparent lattice curvature, very high densities of them are required to fit noise in the measured orientation gradients. This effect results in grains with certain orientations having very high noise floors for densities of one or more types of GND.

A second reason that GND density estimates generally provide a lower bound on the total dislocation density is that some dislocation arrangements generate no net orientation gradient between measurement points. These dislocations are termed statistically stored dislocations (SSDs) (Arsenlis and Parks, 1999). A simple example is the presence of a dislocation dipole between EBSD measurement points. The opposite senses of lattice curvature of the two dislocations cause their lattice rotations to cancel over distances on the order of the spacing between the dislocations. The same effects can occur with more complex dislocation arrangements, including subgrain boundaries. If two subgrain boundaries generate lattice curvatures of equal magnitude but opposite sense between measurement points then their rotations cancel and are not detected in the orientation data. In such a dataset, the dislocations that are not detected are SSDs. These effects are illustrated schematically in Figure 3, though it should be noted that it is better practice to separate the dislocation density into GND and SSD contributions rather than assigning individual defects. Clearly, the fractions of the dislocation population that appear as GNDs or SSDs are not fixed but depend on the mapping step size, dislocation arrangement, and positions of the measurement points relative to the dislocations (Jiang et al., 2013a; Ruggles et al., 2016b; Wallis et al., 2016). In general, as step size is increased, a greater fraction of the dislocation population will become SSDs since the lattice curvature that they generate is more likely to be cancelled by dislocations of the opposite sign.

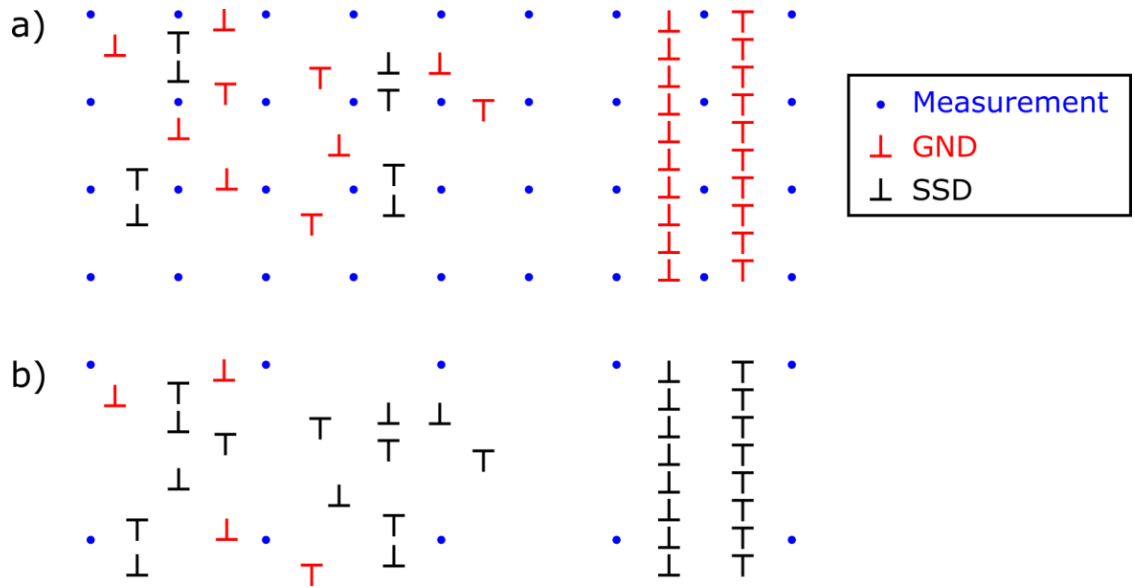

**Figure 3**. Schematic illustration of the effect of mapping step size on whether dislocations are evident as geometrically necessary dislocations (GNDs) or are statistically stored dislocations (SSDs). The two arrays of dislocations are identical in (a) and (b), with the only difference being the grid spacing of orientation measurements indicated as blue points. In (a) most dislocations appear as GNDs. In (b) the same dislocation arrangement is analysed with a step size double that in (a), causing most dislocations to act instead as SSDs. Formally, we do not know which individual dislocation is a GND, but we do know that their net contribution must sum to the total shown in this schematic.

## 4. Practical Implementation

The surface preparation of specimens for HR-EBSD must be of the highest quality. With strain sensitivity on the order of $10^{-4}$, the technique is easily capable of detecting damage remaining in the specimen surface from cutting and grinding. As shifts in ROIs are measured with subpixel accuracy, any scratches visible in conventional EBSD maps based on diffraction pattern quality (e.g., band contrast, band slope, image quality) will also be evident in the resulting HR-EBSD maps. Even very fine scratches evident in forescattered electron images during map setup are likely to be evident in the final HR-EBSD results. We typically prepare geological minerals using a standard polishing routine of progressively finer diamond suspensions, finishing with either 0.05 μm diamond or colloidal silica (Lloyd, 1987), on thin, hard polishing cloths to minimise topography. However, for HR-EBSD we take the extra precaution of cleaning polishing cloths, samples, and sample holders in an ultrasonic water bath between each polishing step.

As HR-EBSD analysis is largely a post-processing technique, most aspects of data acquisition are shared with conventional EBSD (Wilkinson and Britton, 2012). Diffraction patterns are collected from a highly polished specimen surface tilted at 70° to the incident beam and are probed with high-energy electrons. Backscattered electrons generate diffraction patterns on a phosphor scintillator screen, which is imaged using a camera based on either a charge-coupled device or complementary metal-oxide-semiconductor sensor. Optimal acquisition systems for HR-EBSD are capable of rapidly collecting and saving megapixel diffraction patterns with minimal optical distortion, and precise camera movement and positioning (Britton et al., 2013a; Maurice et al., 2013).

A few additional data acquisition steps are necessary to obtain all the information required for HR-EBSD processing. Shifts in the position of the diffraction pattern due to scanning of the beam across the specimen surface require that a correction is applied to the position of the pattern centre (the point on the phosphor screen closest to the source of the diffraction pattern on the specimen surface, Figure 1) (Wilkinson et al., 2006a). This correction is calculated from data obtained by scanning an undeformed single-crystal standard (typically a Si wafer), in which pattern shifts are guaranteed to result only from beam scanning and not distortion of the crystal lattice. To apply this correction and the pattern remapping procedure, the position of the pattern centre must be known as accurately and precisely as possible. Some EBSD systems include an automated routine to determine the pattern centre based on collecting patterns over a range of camera insertion distances (Maurice et al., 2011), which can be applied before acquisition of each dataset to determine the pattern centre specific to the precise specimen-camera geometry used for that experiment. Reference frame conventions used in all aspects of data acquisition and processing should be validated using the approach of Britton et al. (2016) or similar.

The optimal settings for acquisition of the diffraction patterns and the map itself depend on the aims of the analysis. Generally, for optimal strain and stress sensitivity, diffraction patterns should be collected with minimal binning and gain. Typically, shifts in ROIs caused by lattice rotations are much greater than those caused by elastic strains. Therefore, if lattice rotations or GND densities are the target of the analysis, then binning of pixels in the diffraction patterns can be applied to increase acquisition speed with negligible impact on the results, provided that the rotations are sufficiently large (Jiang et al., 2013a; Wallis et al., 2016). Map areas should be small enough that movement of the pattern centre during beam scanning can be accurately corrected (Wilkinson et al., 2006a), which in practice typically limits map dimensions to a few hundred micrometres. Two main considerations, alongside the length scales of the microstructure, affect the choice of step size. First, the minimum orientation gradient and hence GND density that can be detected is inversely proportional to the step size (Equation 15). This consideration does not apply to the elastic strain and stress measurements as they are not based on spatial gradients. Second, smaller step sizes result in larger data storage requirements and longer times for acquisition and post-processing. Typical datasets, a few hundred points in each map dimension with little or no binning of the diffraction patterns, generally require tens to hundreds of gigabytes of storage and a few days of processing on a desktop workstation. A powerful approach is to collect datasets at the highest practical spatial resolution and subsequently undersample the data points in post processing to systematically investigate the effect of increasing step size on GND density estimates (Jiang et al., 2013a; Ruggles et al., 2016b; Wallis et al., 2016). Images of diffraction patterns must be saved at the maximum possible bit depth for HR-EBSD post-processing (Britton et al., 2013a).

Cross-correlation analysis takes place offline and works on image files of the diffraction patterns exported from the acquisition software. The analysis is typically performed in software with image processing and matrix algebra capabilities, such as MATLAB®, or dedicated HR-EBSD packages, such as CrossCourt 4 (BLG Vantage). Requirements for running the analysis include knowledge of the pattern centre position and its correction for beam scanning, knowledge of the elastic constants (for calculation of stresses and separation of normal strains) and possible dislocation types (for calculation of GND densities), selection of reference points, and choice of the size, number, and positions of the ROIs. Reference points are generally chosen within regions of high pattern quality and, where possible, that are likely to be under minimal elastic strain, as measured elastic strains are relative to the strain state of the reference point. In deformed geological materials the strain state at the reference point is typically unknown and it is likely that there are few or no areas that are free from elastic strain. However, strains and stresses can be readily recalculated relative to the mean strain and stress state within each grain area, which is a more intuitive form to interpret and is independent of the choice of reference point (Jiang et al., 2013b; Mikami et al., 2015; Wallis et al., 2017a). Similarly, GND density estimates are independent of the choice of reference point as they are calculated from the spatial gradients of the rotation fields.

## 5. Example HR-EBSD datasets

### 5.1. Data acquisition and processing

In this section, we present datasets that illustrate several of the main points and considerations for HR-EBSD analyses in general and geological minerals in particular. The data were acquired on an FEI Quanta 650 field emission gun SEM equipped with an Oxford Instruments AZtec EBSD system and NordlysNano EBSD detector in the Department of Earth Sciences, University of Oxford. Reference frames for data acquisition and processing were validated following the approach of Britton et al. (2016). The pattern centre was determined prior to each run using an automated camera stepping routine in the acquisition software, implementing a process similar to that proposed by Maurice et al. (2011). Shifts in the pattern centre due to beam scanning were calibrated on an undeformed single crystal Si standard (Wilkinson et al., 2006a; Wallis et al., 2016). All datasets were collected at the full resolution of the EBSD detector giving diffraction patterns of 1344 x 1024 pixels. All datasets were processed using 100 ROIs of 256 x 256 pixels and the robust iterative fitting and pattern remapping approaches of Britton and Wilkinson (2011, 2012b). Data points were filtered out if they had either a mean angular error > 0.004 radians in the deformation gradient tensor or a normalised peak height < 0.3 in the cross-correlation function (Britton and Wilkinson, 2011). Details of the datasets are presented in Table 1. In the sections that follow, we highlight aspects of the results that are particularly relevant to the HR-EBSD method and direct interested readers elsewhere for detailed discussions of the samples and geological implications of the results.

**Table 1**

| Dataset | Figure (s) | Sample number | Associated publication | Map size (data points) | Step size (µm) |
|---|---|---|---|---|---|
| Nanoindent in olivine | 4 and 5 | MN1 | (Kumamoto et al., 2017) | 258 x 185 | 0.2 |
| Olivine aggregate | 6 and 8 | PI-1523 | (Hansen et al., 2011) | 172 x 116 | 0.2 |
| Olivine single crystal | 7 | PI-1436 | (Wallis et al., 2017a) | 450 x 320 | 1.0 |
| Chessboard subgrains in quartz | 9 | P13052 | (Wallis et al., 2017b) | 92 x 67 | 8.0 |

*5.2. Lattice rotations and absolute elastic strain/residual stress heterogeneities*

Figure 4 compares the precision in misorientation angles obtained by conventional EBSD and HR-EBSD processing of the same set of diffraction patterns. This dataset was acquired from a nanoindent made with a triangular pyramidal (Berkovich) diamond tip in a single crystal of San Carlos olivine as part of the study by Kumamoto et al. (2017). This dataset has the advantage that the olivine single crystal was well annealed with dislocation densities on the order of $< 10^{10}$ m$^{-2}$ prior to indentation (Wallis et al., 2016; Kumamoto et al., 2017). Therefore, regions of the map outside the zone of deformation around the indent are suitable for analysing the noise levels of the measurement techniques. The maps of local misorientation, calculated as the average misorientation angle within a 3 x 3 pixel kernel centred on each measurement point, immediately highlight the difference in noise level between the conventional EBSD and HR-EBSD data. This difference is quantified further in profile A–A', which presents point-to-point misorientation angles far from the indent. Two standard deviations of misorientation angles in the conventional EBSD data is 0.12°, which is reduced to 0.02° in the HR-EBSD data. The difference that this improved precision makes to the ability to resolve subtle orientation gradients is apparent in profile B–B', which presents misorientation angles relative to the orientation at point B, in a transect across the indent. The HR-EBSD data clearly resolve the subtle structure better than the conventional EBSD data, particularly at distances of 35–40 µm where the orientation gradient is largely obscured by noise in the conventional EBSD data.

Figure 5 presents the distributions of elastic strain and residual stress around the same nanoindent. In this case, the indent dataset makes a good example of a dataset in which absolute values of strain and stress can be obtained by HR-EBSD because the crystal lattice at the reference point, chosen to be far from the indent, should be essentially unstrained. This assumption is supported by the uniformity of the strain fields outside the zone of influence of the indent. These far-field regions also demonstrate precision in strain and stress measurements on the order of $10^{-4}$ and a few tens of megapascals, respectively. The technique clearly resolves the strain and stress fields around the indent and linear microcracks extending from it, with magnitudes of the in-plane compressive normal stress locally exceeding 1 GPa.

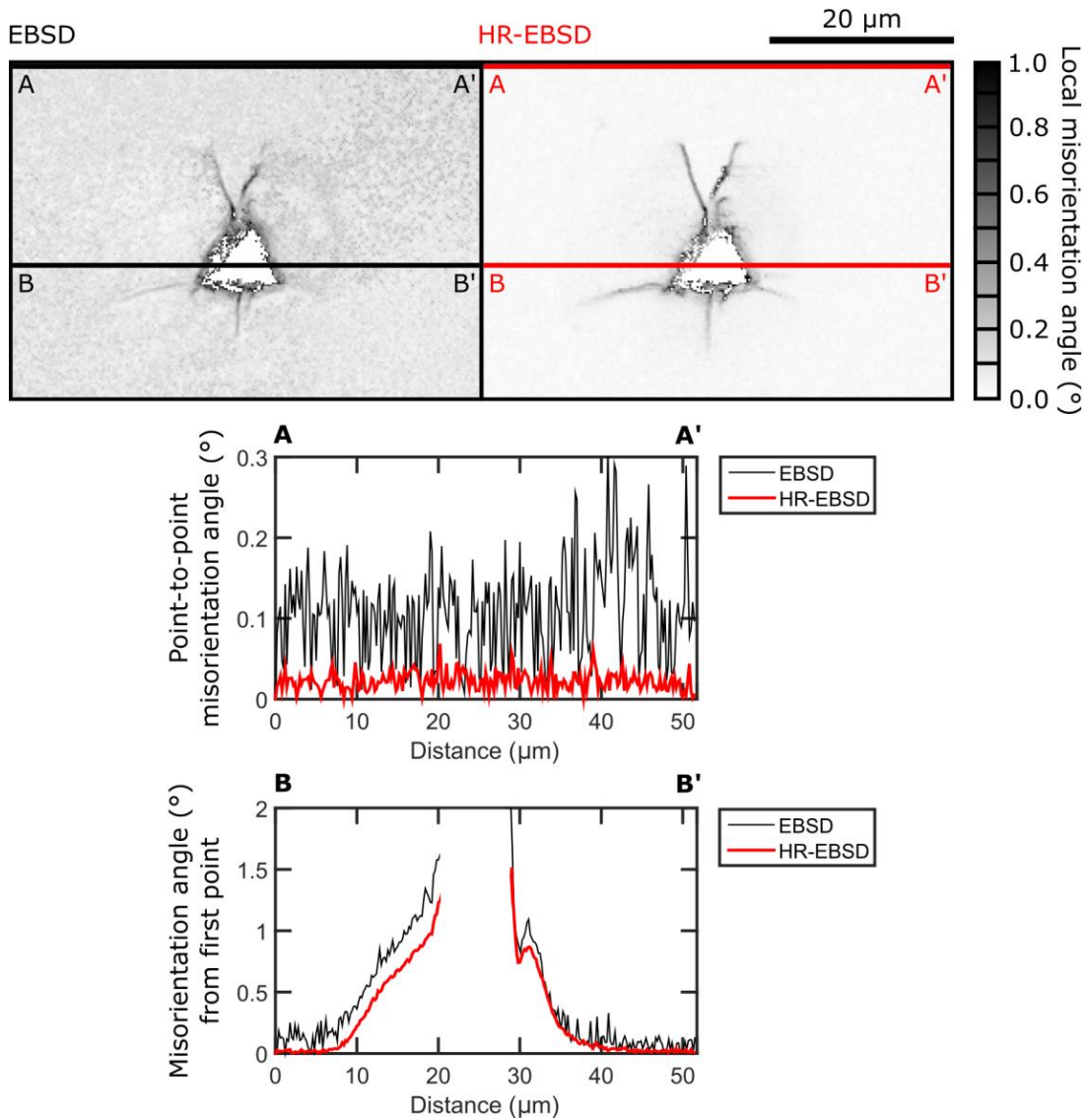

**Figure 4**. Misorientation data generated by conventional EBSD and HR-EBSD processing of the same diffraction patterns collected across a Berkovich nanoindent in an olivine single crystal. Local misorientation maps present the average misorientation between pixels in a 3x3 pixel kernel centred on each measurement point. Profile A–A' presents point-to-point misorientation angles far from the indent and therefore demonstrates the noise in the measurement methods. Profile B–B' presents misorientation angles relative to the orientation at point B across the indent and demonstrates the ability of each measurement method to reveal orientation gradients around the indent.

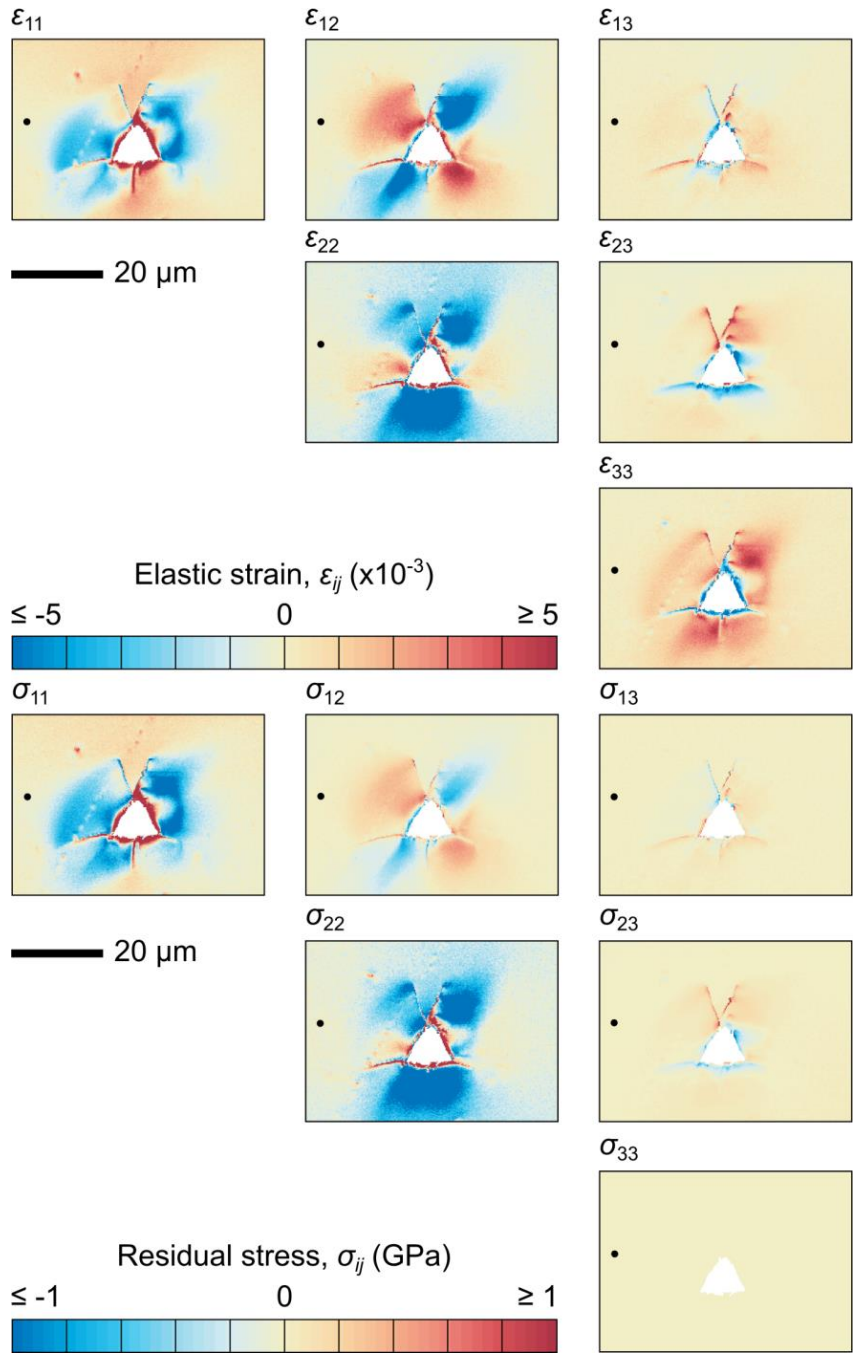

**Figure 5**. Maps of elastic strain ($\varepsilon_{ij}$) and residual stress ($\sigma_{ij}$) around the same indent as in Figure 4. All data are relative to the strain and stress state at the reference point marked in black. The $\sigma_{33}$ component of the stress tensor is constrained to be zero to enable calculation of the $\varepsilon_{33}$ component of the strain tensor (Britton and Wilkinson, 2012b). Tensional stresses are indicated by positive values and compressional stresses are indicated by negative values.

*5.3. Pattern remapping and relative elastic strain/residual stress heterogeneities*

Figure 6 displays the effect of data processing procedures applied to datasets acquired from microstructures containing lattice rotations of several degrees, which are typical of crystalline aggregates deformed at high temperatures. This dataset was obtained from an aggregate of San Carlos olivine (sample PI-1523) shortened to 17% plastic strain at temperatures in the range 1373–1523 K by Hansen et al. (2011), imparting intragranular misorientations of several degrees. In this case, it is likely that most of the material is subject to some elastic strain, as indicated by the continuously varying strain distributions. Therefore, the calculated strains, and hence stresses, are not absolute values but are relative to the unknown strain states at the reference points. Figure 6a presents strains measured after a single pass of cross correlation, whereas Figure 6b presents strains recalculated following remapping of the test patterns into the orientations of the reference patterns and a second pass of cross correlation. Figure 6c presents the difference between these datasets and reveals that erroneous strains on the order of $10^{-3}$ to $10^{-2}$ were removed by the remapping procedure. This result is consistent with the work of Jiang et al. (2013b) on polycrystalline copper deformed to plastic strains of several percent. Figure 6d demonstrates that the intragranular stress heterogeneities calculated from the strains after pattern remapping and the second pass of cross correlation generally still have magnitudes on the order of 1 GPa. In Figure 6e, these stresses have been recalculated by subtracting the mean value within each grain area to remove the effect of the choice of reference points (Jiang et al., 2013b; Mikami et al., 2015; Wallis et al., 2017a). The result gives the intragranular stress heterogeneities relative to the (unknown) mean stress state within each grain area.

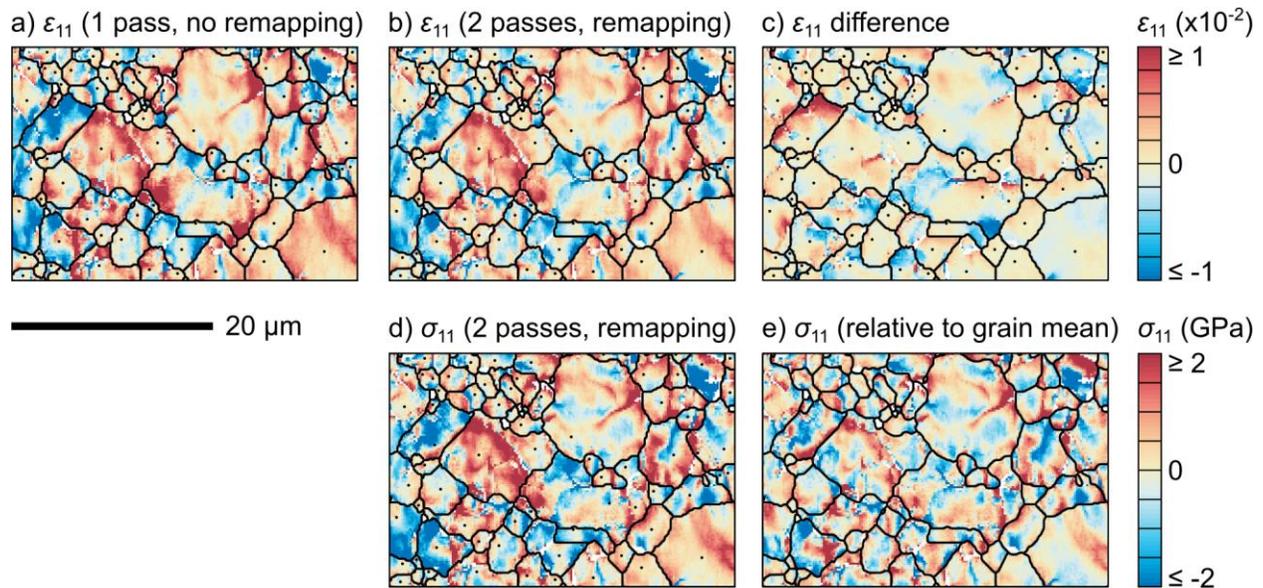

**Figure 6**. The effect of data processing procedures on measured elastic strains and residual stresses. The top row displays the $\varepsilon_{11}$ component of the strain tensor (a) after one pass of cross correlation, (b) after two passes of cross correlation incorporating remapping of the test patterns into the orientations of the reference patterns, and (c) the difference between (a) and (b). Black lines indicate grain boundaries with $\geq 10°$ misorientation. Black dots indicate the reference point for each grain. The bottom row displays the $\sigma_{11}$ component of the stress tensor (d) computed after two passes of cross correlation and pattern remapping (i.e., corresponding to the strains in (b)) and (e) recalculated relative to its mean value within each grain.

### 5.4. Geometrically necessary dislocations

*5.4.1. Comparison between GND densities from conventional EBSD and HR-EBSD*

Figure 7 presents a comparison of GND densities in a single crystal of olivine calculated from orientation gradients derived from conventional EBSD and HR-EBSD analysis of the same diffraction patterns. The crystal was experimentally deformed at 1473 K with the compression direction running vertically in the maps. Wallis et al. (2017a) discussed the HR-EBSD results in detail, whereas here we focus on their relationship to the conventional EBSD data. GND densities were estimated using the $L_2$ method (Equation 12) of Wallis et al. (2016). In both datasets, the most obvious set of structures are bands of elevated GND density trending top-right to bottom-left. Less prominent arrays of linear features trending top-left to bottom-right result from small rotations across microcracks (Wallis et al., 2017a). Two key differences are evident between the results obtained from conventional EBSD and HR-EBSD. First, the noise level is higher in the conventional EBSD data, particularly on the left side of the map where the details of the GND structures are partially obscured. This contrast is an obvious outcome of the different angular resolutions of the two methods. Second, the dislocation types that the GND densities are assigned to differ between the two datasets. Densities of [100] screw, [001] screw, and (010)[001] edge dislocations are higher in the results from conventional EBSD than in the results from HR-EBSD, whilst the opposite is true of the density of (001)[100] edge dislocations. This difference is a consequence of the different misorientation axes determined by the two methods, with those from HR-EBSD being the more precise (Wilkinson, 2001).

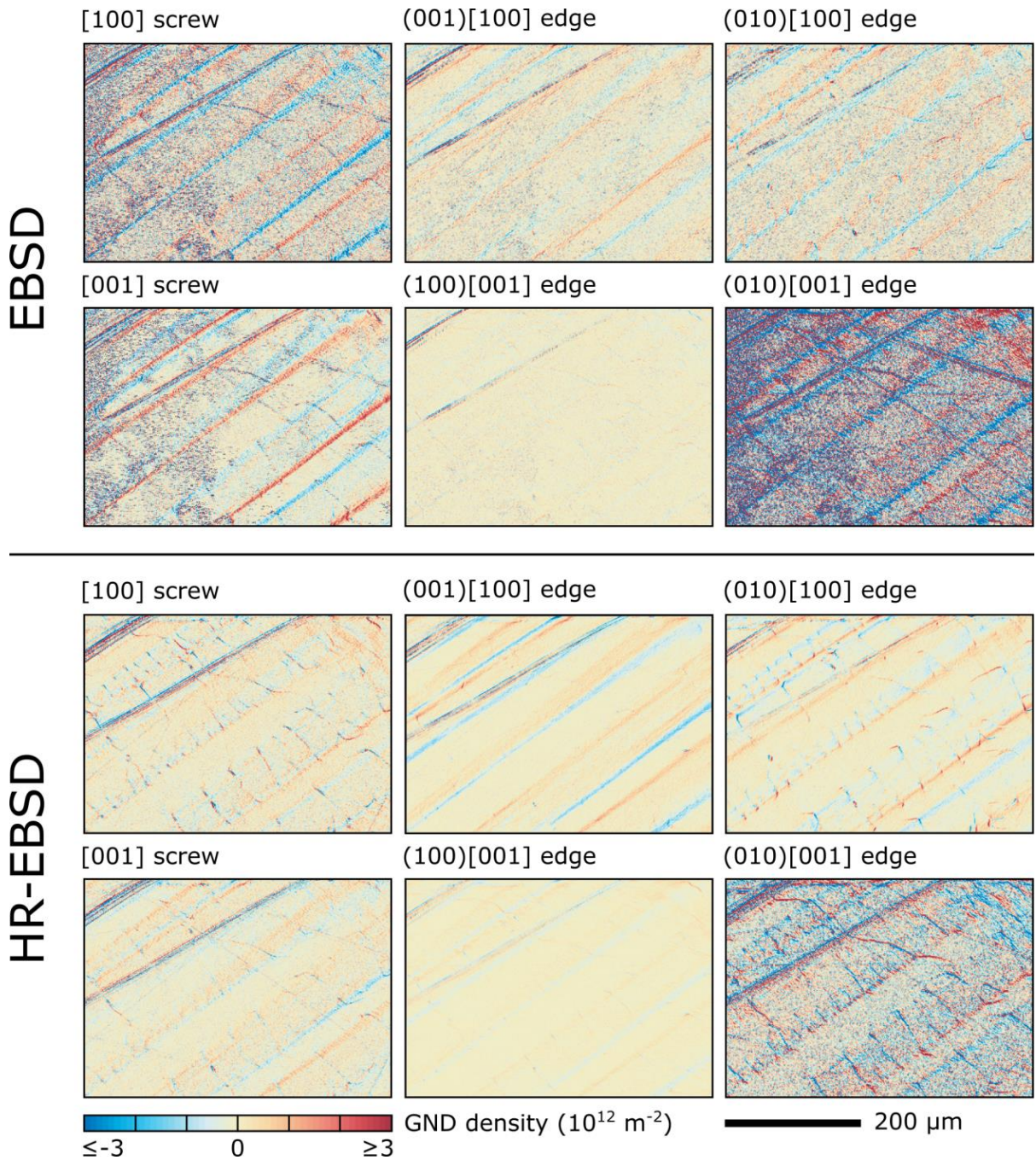

**Figure 7**. Densities of six types of geometrically necessary dislocation in a single crystal of olivine calculated from lattice orientation gradients obtained by conventional EBSD (top) and HR-EBSD processing of the same diffraction patterns (bottom). Positive and negative GND densities indicate dislocations with opposite senses of Burgers vector (i.e., generating opposite senses of lattice curvature).

*5.4.2. The effect of crystal orientation on noise floors in GND density estimates*

Figure 8 illustrates the effect of crystal orientation on the noise floors in GND density estimates using the same dataset as in Figure 6. This aggregate of olivine contains grains in a variety of orientations. Figure 8a colour codes these orientations according to the crystal direction oriented normal to the specimen surface. Figure 8b presents estimates of the noise floor for densities of (010)[100] edge dislocations in these differently oriented grains. To estimate the noise floor, we estimated the precision in orientation gradients ($\phi$) from the angular resolution of the HR-EBSD measurements ($\theta$) and the step size ($d$) by

$$\phi = \frac{\theta}{d}. \tag{16}$$

We estimated $\theta$ to be $3\times10^{-4}$ rad for this dataset based on the results of Wilkinson *et al.* (2006a) and Wallis *et al.* (2016), the diffraction pattern size in Table 1, and comparison of the estimated noise floors to Figure 8c. To predict the GND density noise floors in each grain, we used the crystal orientations of each reference point along with the estimate of $\phi$, instead of measured orientation gradients, as inputs for the same $L_2$ minimisation procedure (Equation 11) applied to the real data. Edge dislocations with [100] Burgers vectors produce little/no detectable lattice curvature when [100] is oriented (sub)normal to the specimen surface, and therefore high densities of them are required to fit the orientation noise (Section 3.2). One example of a grain in this orientation is marked with a black star in Figure 8a–c. Grains with [100] axes at lower angles to the specimen surface generally have lower estimated noise floors for densities of (010)[100] edge dislocations. The measured densities of (010)[100] edge dislocations are presented in Figure 8c. Grains, including the marked example, with high estimated noise floors in Figure 8b generally also exhibit high GND densities in Figure 8c, obscuring any GND structures. Most other grains in this specimen exhibit distinct GND structures, including patches and bands of elevated GND density, resolvable above the noise floor.

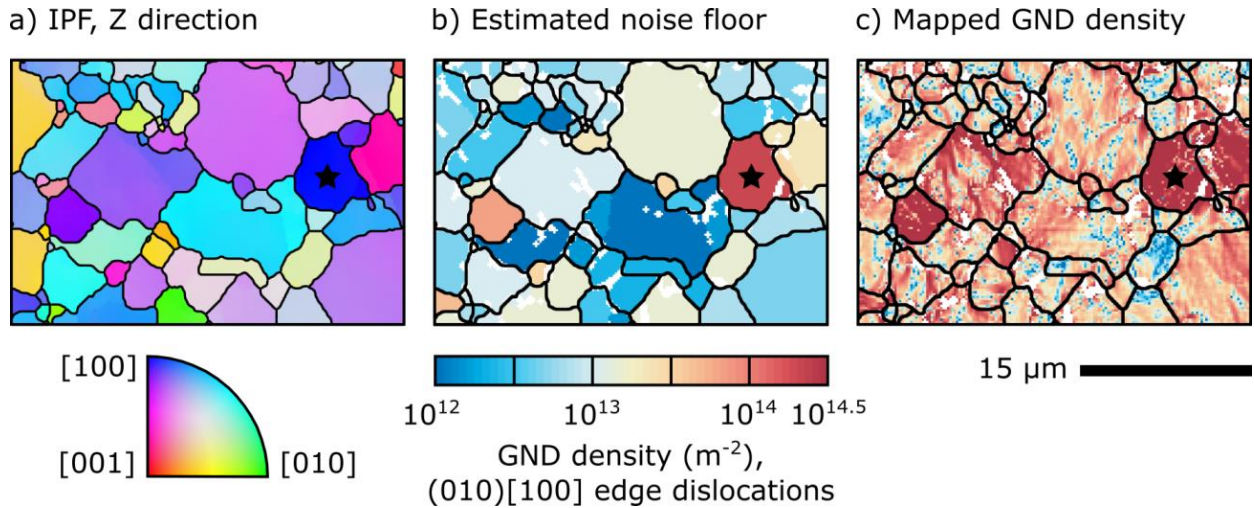

**Figure 8**. The effect of crystal orientation on the noise floor of geometrically necessary dislocation (GND) density estimates. a) Map colour coded for crystal orientation according to the inverse pole figure (IPF) for the Z direction of the specimen (i.e., the normal to the specimen surface). b) Estimated noise floor for densities of (010)[100] edge GNDs. c) Mapped densities of (010)[100] edge GNDs. The black star marks a grain with [100] oriented normal to the specimen surface. Black lines indicate grain boundaries with misorientation angles ≥ 10°.

*5.4.3. Minerals with more than six dislocation types*

In the examples above, dislocation types could be separated using the $L_2$ approach (Section 3.2) because olivine has relatively few slip systems active at high temperature, and therefore considering six dislocation types provides a reasonable characterisation and a unique solution. However, many geological minerals, particularly those with higher symmetry, have many more dislocation types that can plausibly be activated (as is typically the situation in metallurgical studies). Important examples include quartz (Linker et al., 1984; Lloyd et al., 1997), calcite (De

Bresser and Spiers, 1997), and garnet (Mainprice et al., 2004). For cubic minerals, in certain applications it may be appropriate to assume a dominant family of slip systems and solve for an optimised density of each of the associated symmetrically equivalent dislocation types, as has been carried out for cubic metals (Wilkinson and Randman, 2010). For body centred cubic garnet, consideration of 16 dislocation types (12 edge and 4 screw) on the {110}<111> slip system could provide a reasonable approximation (Mainprice et al., 2004). This approach was taken by Wilkinson and Randman (2010) in their analysis of the GND content of body centred cubic Fe, in which they selected a solution for the densities of each dislocation type by employing the $L_1$ scheme to minimise the total line energy (Equation 13). In geological examples of such analyses, the sum of the densities of the symmetrically equivalent dislocation types will often be the result of interest, and their subdivision may be less important.

The situation is more complex for minerals, such as trigonal quartz and calcite, with more than six dislocation types that are spread across multiple families of slip systems. In such cases, typically the family of slip systems cannot be assumed *a priori* and instead is the information of interest. Moreover, the abundance of dislocation types generates considerable redundancy in solving Equation 10, and again the $L_1$ scheme must be employed to minimise some other variable, such as the total line energy, but now the specific dislocation types favoured by the minimisation are of importance. This approach has been employed for hexagonal close packed metals, such as Ti (Britton et al., 2010; Britton and Wilkinson, 2012a). Wallis et al. (2017b) adopted this method in analyses of chessboard subgrain boundaries in quartz, in which they considered the six families of dislocation types with either <*a*> or [*c*] Burgers vectors presented in Figure 9. Transmission electron microscope and visible light microscope observations of chessboard subgrain boundaries indicate that they are composed primarily of {*m*}[*c*] and (*c*)<*a*> edge dislocations (Blumenfeld et al., 1986; Mainprice et al., 1986; Okudaira et al., 1995; Kruhl, 1996). In the HR-EBSD results in Figure 9 and the other samples analysed by Wallis et al. (2017b), {*m*}[*c*] edge dislocations are abundant, but (*c*)<*a*> edge dislocations are largely absent. Instead, <*a*> screw dislocations are apparent in high densities, particularly in boundaries with traces parallel to [0001], which would otherwise be expected to be composed of (*c*)<*a*> edge dislocations. Wallis et al. (2017b) attributed the difference between the HR-EBSD results and previous results to the lower energy of screw dislocations relative to edge dislocations, which results in them being favoured in the energy minimisation scheme. Wilkinson and Randman (2010) also noted that the $L_1$ energy minimisation scheme returned greater densities of screw dislocations than edge dislocations in their analysis of Fe. Despite these complications, Wallis et al. (2017b) highlighted that it is possible to unambiguously discriminate dislocations with <*a*> Burgers vectors from those with [*c*] Burgers vectors using this approach.

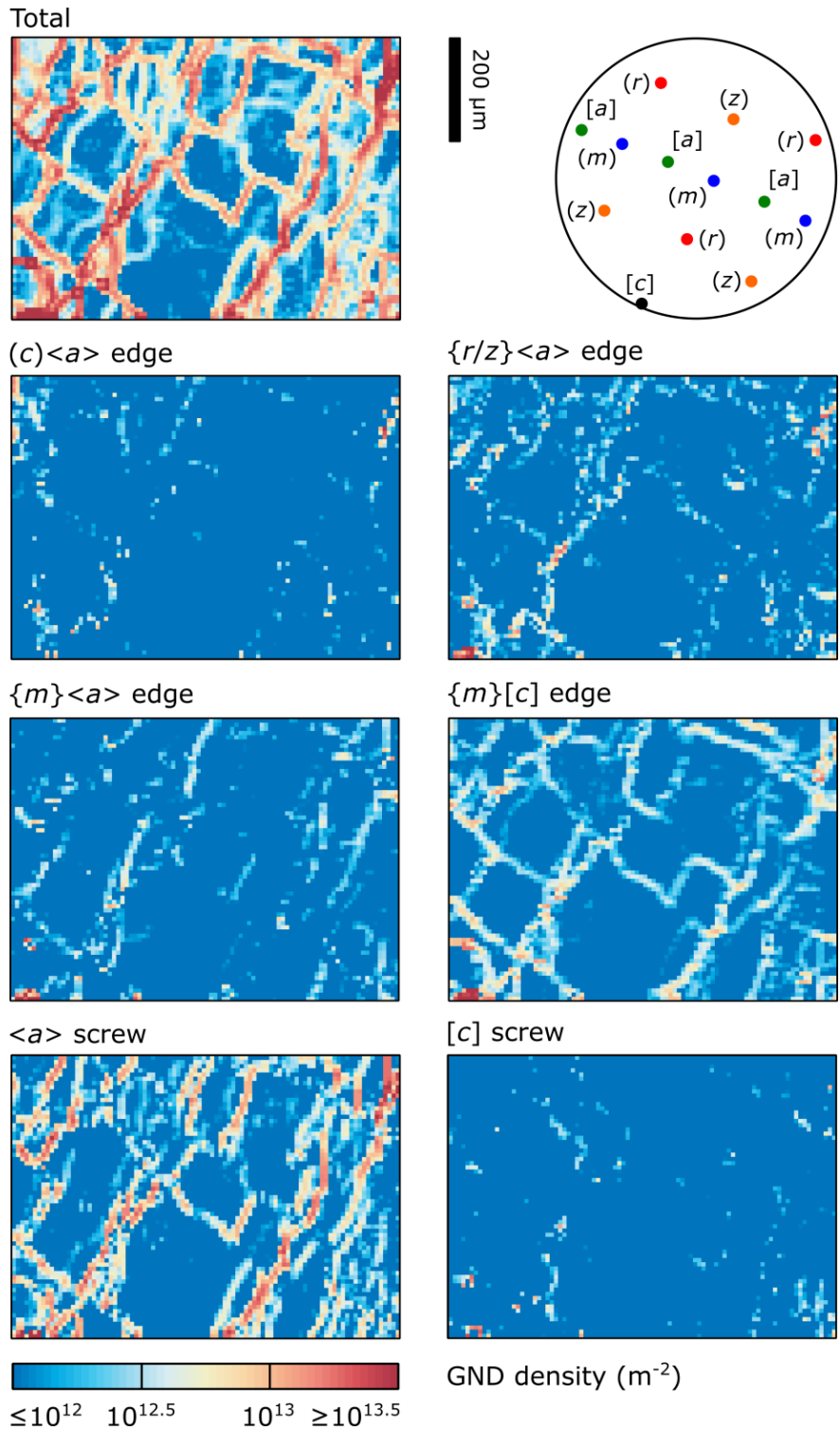

**Figure 9**. GND densities in chessboard subgrain boundaries in quartz from the Greater Himalayan Sequence, Nepal. The pole figure indicates the crystal orientation. From Wallis et al. (2017b).

## 6. Strengths of HR-EBSD for analysing intracrystalline lattice distortion

HR-EBSD has several obvious advantages over conventional, Hough transform-based EBSD for analysis of intragranular lattice distortions. These advantages stem from the improved precision in misorientation angles and axes, along with the ability to map heterogeneities in elastic strain and residual stress (Wilkinson, 1996, 2001; Wilkinson et al., 2006a; Britton and Wilkinson, 2012b). These capabilities have been widely exploited in the materials sciences (Britton et al., 2010; Littlewood et al., 2011; Britton and Wilkinson, 2012a; Villanova et al., 2012; Wilkinson and Britton, 2012; Maurice et al., 2013; Jiang et al., 2013b, 2013c, 2015a, 2016; Vilalta-Clemente et al., 2017). However, geological materials present a diverse new array of crystal structures, microstructures, deformation mechanisms, and conditions of formation and deformation, that can be investigated using HR-EBSD. Therefore, in this section we discuss in a general sense the benefits that HR-EBSD brings to analysis of geological minerals in particular.

Many deformation microstructures relevant to geological interpretations involve very small misorientation angles. Some of these microstructures, including deformation lamellae (Trepmann and Stöckhert, 2003), slip bands (De Bresser, 2002), and undulose extinction (Halfpenny et al., 2006), are the subtle expressions of limited dislocation activity at low homologous temperatures. Other common microstructures form at high homologous temperatures, at which differential stresses and hence dislocation densities are often low, and include incipient subgrain boundaries (Lloyd et al., 1997; Wheeler et al., 2009; Wallis et al., 2017b) and low densities of free dislocations in subgrain interiors (De Bresser, 1996; Qi et al., 2018). Analyses of these microstructures clearly benefit from the precision in misorientation angles offered by HR-EBSD. This effect is demonstrated in Figures 4 and 7, where the subtleties of microstructures formed at low and high temperatures, respectively, are revealed in new detail by HR-EBSD processing of the diffraction data. Moreover, the improved misorientation axes provided by HR-EBSD (Wilkinson, 2001) make it possible to reliably investigate the components of lattice curvature and hence types of dislocations contributing to these microstructures. The full benefits of improved misorientation axes are realised in applications to geological minerals due to the diversity of dislocation types and associated slip systems that they exhibit (e.g., Figures 7 and 9). Deformation of high-symmetry crystal structures, such as the cubic metals, is generally accommodated primarily by one family of symmetrically-equivalent dislocation types (Wilkinson and Randman, 2010; Jiang et al., 2013c), reducing the need to discriminate between them. The generally lower symmetries of geological materials typically require that more than one family of dislocation types are required to be active to accommodate an arbitrary deformation (Morales et al., 2014; Detrez et al., 2015), and their associated slip systems typically have different strengths (Linker et al., 1984; Bai et al., 1991; De Bresser and Spiers, 1997). Therefore, determining the dislocation types and slip systems that were active is often one of the principal goals of geological studies. These analyses require precise misorientation axes to correctly populate the Nye tensor, $\boldsymbol{\alpha}$, in Equation 8 and $\boldsymbol{\lambda}$ in Equation 10. This benefit of HR-EBSD processing is highlighted in Figure 7, in which the dominant types of dislocation forming the substructure differ between the conventional EBSD and HR-EBSD data. Complementary approaches to analysing the dislocation content based on the Nye tensor, such as the weighted Burgers vector of Wheeler *et al.* (2009), could benefit similarly from taking HR-EBSD misorientation data as inputs.

Alongside generating lattice misorientations, the diverse and complex histories of many geological materials involve many potential sources of heterogeneous residual stress and elastic strain (Friedman, 1972; Holzhausen and Johnson, 1979). Structural defects in rocks are one source of residual stress heterogeneity and include dislocations (Anderson et al., 2017) and microcracks (Sun and Jin, 2012). Another diverse group of sources encompasses residual stresses arising from interactions among grains with anisotropic properties. An aggregate of grains that are elastically and plastically anisotropic can have stresses locked in during deformation as the shape changes of the grains prevent full relaxation upon removal of the macroscopic applied load (Friedman, 1972). Rocks can accrue additional residual-stress heterogeneity during decompression and cooling due to anisotropic elastic properties and thermal expansivities, respectively (Rosenfeld and Chase, 1961; Holzhausen and Johnson, 1979). Similarly, phase changes potentially introduce stress heterogeneity as a result of changes in volume, which drive additional processes, such as transformation plasticity (Poirier, 1982). HR-EBSD can also map changes in the relative lengths of unit cell axes resulting from solid solutions (Schäffer et al., 2014; Speller et al., 2014). These compositional variations manifest as pseudo-strain heterogeneity due to variations in interplanar angles.

Interpretation of the geological processes recorded in residual stress fields has been hindered by the challenge of measuring elastic strains and residual stresses at the grain scale. Techniques that are currently employed include Raman spectroscopy (Kohn, 2014) and X-ray Laue microdiffraction (Chen et al., 2015, 2016; Boullier et al., 2017). Another technique that could potentially be employed is convergent beam electron diffraction in the transmission electron microscope (Champness, 1987). HR-EBSD has advantages over each of these techniques that make it a particularly appealing option. Whilst Raman spectroscopy can be applied to unrelaxed grains confined within a specimen volume, it provides only a scalar measure of confining pressure, rather than the full stress tensor. Absolute values of the full stress tensor can be obtained from X-ray Laue microdiffraction, but such measurements require access to a synchrotron X-ray source, making routine analysis of large sample sets challenging. In contrast, HR-EBSD provides a means to obtain broadly comparable data using a standard scanning electron microscope. Moreover, HR-EBSD offers improved spatial resolution (potentially < 100 nm) relative to Raman spectroscopy and X-ray diffraction (Dingley et al., 2010). Transmission electron microscopy provides even greater spatial resolution, but at the expense of areal extent and at the risk of greater modification of the stress state during specimen preparation (Dingley et al., 2010). HR-EBSD provides a combination of precision, spatial resolution, and areal coverage that are well suited to the microstructures of deformed rocks with characteristic length-scales in the range $10^{-7}$ to $10^{-4}$ m.

As with all microstructural data, the interpretation of HR-EBSD results is not trivial. Mapped GND density and stress fields record the net effect of potentially complex deformation histories that may include one or more deformation events of interest as well as additional processes, such as generation of stresses during decompression and cooling. These challenges apply to samples deformed in both nature and experiments. However, these challenges are not specific to HR-EBSD data but apply also to microstructural data acquired by any technique. Rather, the improved level of microstructural detail provided by the precision of HR-EBSD can only aid in the interpretation of processes that controlled microstructural evolution during the history of the rock. For example, the ability to resolve dislocation structures associated with subtle misorientations, to provide new constraints on the dislocation types that contribute to them, and assess their association with stress heterogeneities helps to constrain the processes of microstructural evolution in both experimental (e.g., Wallis et al., 2017a) and natural (e.g., Wallis et al., 2017b) samples. Ongoing and future work will focus on developing strategies for deciphering the contributions of different causes to heterogeneous residual stress fields.

## 7. Current limitations of HR-EBSD and areas for further development

Several caveats should be borne in mind when designing HR-EBSD experiments and interpreting HR-EBSD data. Some of these considerations stem from the data processing procedures and are therefore specific to HR-EBSD. However, several key points are inherent to (mis)orientation and stress datasets more generally, regardless of the technique used to acquire the data.

The characteristics of a microstructure can place constraints on the information that can be revealed by HR-EBSD analysis. A limitation of current data processing procedures is that crystal orientations at a given pixel must be within 11° of the orientation of the reference point (Britton and Wilkinson, 2011, 2012b). Beyond this range, the patterns to be compared are too dissimilar and distorted for reliable cross-correlation analysis. This constraint limits the maximum intragranular orientation range that can be analysed with one reference point to 22°, assuming that a reference point with the optimal orientation can be found in advance. Grains within rocks deformed to large plastic strains at high homologous temperatures commonly develop arrays of subgrains that can result in intragranular orientation ranges > 22° (e.g. Cross et al. (2017)). In such cases, more than one reference point must be chosen (e.g., one reference point per subgrain) and stress states will generally not be directly comparable between the areas associated with each reference point. It is possible that this limitation could be overcome by developing a routine for cross correlating between two or more reference points chosen to be within 11° of one another. For example, if the stress states of two points with orientations differing by 10° were cross-referenced by the cross-correlation procedure, and both those points were subsequently used as reference points, then the orientation range could be extended to 32°. However, this procedure has not yet been applied in practice.

The orientation in which a specimen is sectioned can limit or optimise the information recovered. Some sections through a crystal are better than others for revealing the lattice curvatures induced by possible dislocation types (Wheeler et al., 2009; Wallis et al., 2016). Poorly oriented sections may not reveal lattice curvature generated by an important dislocation type and may result in high noise floors in estimates of GND density, as illustrated in Figure 8. Importantly, this caveat applies not only to HR-EBSD data but also to conventional EBSD and all other orientation data collected on two-dimensional sections. Fortunately, often specimens can be deliberately sectioned in an optimal orientation for revealing the dislocation content based on *a priori* knowledge of the crystal orientation or interpretation of the likely crystallographic preferred orientation. This approach has been applied to single crystals (Wallis et al., 2016, 2017a) and aggregates (Qi et al., 2018) of olivine, which can be sectioned such that the [100] and [001] Burgers vectors generally lie at low angles to the plane of the section. Similar considerations apply to analysis of stress heterogeneity as the stress state is modified by sectioning of the sample. In particular, the normal stress on the surface is relaxed, whilst the other stress components are modified to a lesser extent due to the Poisson effect and changes in the tractions on the surface (Pagliaro et al., 2011; Kartal et al., 2015). However, if a specific component of the stress state is of particular interest (e.g., the shear stresses acting on a slip system) then a section orientation can be chosen that minimises the extent to which that component is modified during sectioning. A complimentary approach, which has not yet been exploited in HR-EBSD analysis of geological materials, would be to analyse mutually perpendicular sections. The combined dataset would provide, at least in a statistical sense, a more complete characterisation of the orientation gradients (i.e., the GND content), along with differential relaxation of each stress component in each section. To date, fully three-dimensional

HR-EBSD, based on serial sectioning, has not been achieved due to the difficulty of aligning the sections with sufficient precision and correcting changes in pattern centre position.

Several approaches to interpreting the types and densities of GNDs from orientation gradients are available, and each is associated with advantages and disadvantages. The method most widely employed in analysis of conventional EBSD data is to plot misorientation axes on an inverse pole figure and to interpret the dislocation types most likely to have generated the misorientations (Lloyd et al., 1997; Prior et al., 2002; Bestmann and Prior, 2003). Whilst this approach is simple and intuitive, it has the drawbacks of being qualitative, being difficult to decipher the combined effects of multiple dislocation types, and relying on (often implicit) assumptions about the possible dislocation types. A method that has the benefits of being quantitative and not relying on assumptions about possible dislocation types is the weighted Burgers vector approach of Wheeler et al. (2009), which employs the fully constrained $a_{i3}$ components of the Nye tensor. However, this analysis does not exploit the additional constraints on the dislocation content that can be gleaned from other components of the Nye tensor if gradients in elastic strain are assumed to be small (as they are in all analyses based on conventional EBSD, in which elastic strains are not measured). In contrast, the approach outlined in Section 3.2 exploits all of the available orientation gradients, potentially providing a more complete description of the dislocation content, but requires assumptions about the possible dislocation types in order to find their best-fit densities. Furthermore, cases in which more than six dislocation types must be considered require an additional assumption, such as dislocations occupying a minimum-energy configuration, to choose a single solution to Equation 10. As the presence of dislocations implies that their energy is not at a minimum, minimising energy in Equation 13 may not always give an appropriate solution, as suggested by the analysis of quartz in Section 5.4.3 and Wallis et al. (2017b). Nonetheless, the method outlined in Section 3.2 does provide a fully quantitative approach that exploits all available components of the orientation gradients, and in which all assumptions are made fully explicit. We emphasise that each of the above approaches to GND density analysis could be applied to orientation data collected by any method and therefore their strengths and limitations are not related to the method by which the orientation data were acquired (e.g., conventional EBSD, HR-EBSD, X-ray diffraction, etc.). Instead, all analyses of intragranular misorientations can benefit from the precise rotation data offered by the cross-correlation approach of HR-EBSD.

A notable limitation of elastic strain and residual stress data from HR-EBSD is that absolute measurements can only be obtained if unstrained material is present to provide a reference point. This condition is met in particular situations, such as the nanoindent in Section 5.2 (Figure 5). However, in most rocks no material can be assumed to be free from elastic strain. Therefore, generally HR-EBSD provides maps of relative heterogeneities in elastic strain and residual stress, as in Section 5.3. Nonetheless, although the strain state of reference points is generally unknown, the data can be normalised relative to the mean value of each strain/stress component in each grain (Figure 6), providing values which are potentially more intuitive to interpret (Mikami et al., 2015). An additional complication is the diversity of potential sources of residual stress in geological materials (Section 6), which makes it challenging to decipher the contributions from particular processes. Fortunately, many geological applications can still benefit from comparisons of relative stress states; for instance, whether stresses are more heterogeneous in different minerals or between different rocks. Such comparisons may also be exploited to constrain causes of stress heterogeneity; for instance, comparing a rock that has undergone deformation and exhumation to one that has only undergone equivalent exhumation.

Recently, some new approaches to HR-EBSD have been proposed, which attempt to simplify the procedure and improve accuracy by using more advanced digital image correlation (DIC) methodologies. Ruggles *et al*. (2018) suggested the use of inverse compositional Gauss Newton DIC to track the changes in shape of ROIs along with shifts in their positions. Similarly, Vermeij and Hoefnagels (2018) have developed a method that uses finite-strain integrated DIC to correlate the full diffraction patterns in one step, circumventing the use of ROIs and pattern remapping. Recently, Vermeij *et al*. (2019) extended this approach to suggest that simultaneous correlation of all overlapping areas in multiple diffraction patterns can, in theory, be exploited to optimise crystal orientation, stress state, and pattern centre coordinates, providing measurements of absolute stress state extending across multiple grains. So far, these approaches appear promising in tests on simulated diffraction patterns but have not been applied to, or rigorously tested on, experimental diffraction patterns.

## 8. Promising targets for HR-EBSD analysis

The new capabilities offered by HR-EBSD make it easy to envision applications in many areas of rock deformation and petrology. The precise characterisation of dislocation content and associated stress fields is ideally suited to applications in high-temperature rock deformation, which has been the focus of most initial investigations (Wallis et al., 2016, 2017a, 2017b; Boneh et al., 2017; Kumamoto et al., 2017; Qi et al., 2018). Specimens deformed in laboratory experiments can be analysed using HR-EBSD to inform models of deformation processes and potentially to identify microstructures diagnostic of particular rheological behaviours. Natural specimens can be subjected to similar analysis to assess the applicability of laboratory-based models. Similar applications will likely be found in investigations of dynamic and static recrystallisation (e.g., Boneh *et al*. (2017)) and palaeopiezometry. A key advantage of HR-EBSD in these efforts is the capability to provide quantitative data on both GND densities and residual stresses over length-scales in the range $10^{-1}$–$10^{2}$ μm. This length-scale is sufficient to span multiple grains in most deformed rocks and is therefore ideal for bridging the scales between transmission electron microscopy, which can image individual dislocations and map strain at higher spatial resolutions, and more representative rock volumes.

HR-EBSD will likely also be useful in studies of deformation at low temperatures. Stress concentrations associated with compaction or fracturing are ideal targets. HR-EBSD has been applied in studies of crack nucleation in nickel-based superalloys (Jiang et al., 2015b; Zhang et al., 2015) and stress concentrations around the tips of microcracks have been observed in olivine single crystals (Wallis et al., 2017a). Deformation at shallow depths and low temperatures should limit the stress heterogeneities resulting from exhumation and cooling, potentially aiding recognition of stress heterogeneities recording prior deformation.

HR-EBSD could also prove useful in petrological studies by revealing intragranular deformation associated with changes in phase and pressure-temperature conditions. Promising targets include deformation around solid and fluid inclusions in mineral grains (Angel et al., 2014; Avadanii et al., 2017) and due to crystal growth or phase transformations (Gardner et al., 2017; van Noort et al., 2017).

## 9. Conclusions

HR-EBSD is a promising technique developed in the materials sciences that has recently been exploited in initial applications to geological materials. The capabilities of the technique make it extremely well suited to analysis of intragranular lattice distortions of deformed minerals. Its key strength is the ability to map lattice rotations and elastic strains with precision on the order of $10^{-}$

[4] and submicron spatial resolution in a scanning electron microscope. These data provide the bases for estimates of GND density and calculation of residual stress heterogeneity. Caveats include the complex factors that influence GND density estimates, the effect of sectioning on stress state, and that maps of stress heterogeneity, rather than absolute stress state, are obtained from most materials. Nonetheless, the depth of information obtained from HR-EBSD promises new insights and advances in many areas of rock deformation and petrology in both laboratory and natural contexts.

**Acknowledgements**

We thank David Kohlstedt and Andrew Parsons for providing samples. We thank Yves Bernabe for his editorial handling of the manuscript and two anonymous reviewers for their constructive comments. D. Wallis, L.N. Hansen, and A.J. Wilkinson acknowledge support from the Natural Environment Research Council grant NE/M0009661. T.B. Britton acknowledges support for his research fellowship from the Royal Academy of Engineering. Data in this paper can be accessed from GFZ Data Services (dataservices.gfz-potsdam.de/portal/).